\documentclass[onecolumn,prd,amsmath,amssymb,floatfix,showpacs]{revtex4}

\usepackage{amsmath,amssymb}

\usepackage{bm}

\usepackage{epsfig}

\usepackage{graphics} 

\usepackage{slashed}

\newcommand{\bpsi}{\overline{\psi}}

\begin{document}

\title{Factorization breaking of four-quark condensates in the Nambu Jona Lasinio model}

\author{Fabio L. Braghin$^1$\footnote{braghin@ufg.br}, Fernando S. Navarra$^2$\footnote{navarra@if.usp.br}}

\affiliation{ $^1$ Instituto de F\'\i sica, Univ. Fed. de Goi\'as, P.B.131, Campus II, 
74001-970, Goi\^ania, GO, Brazil
\\
$^2$ Instituto de F\'\i sica, Univ. de S\~ao Paulo, Rua do Mat\~ao, Travessa R, 187, CEP 05508-090, S\~ao Paulo, SP, Brazil}

\begin{abstract}
We investigate the factorization hypothesis of the four-quark condensate 
$\langle q \bar{q} q \bar{q} \rangle  = \, A \, \langle q \bar{q} \rangle^2$ 
by considering  the   Nambu Jona-Lasinio Model supplemented with sixth and eighth order interactions.
For that purpose we use the bosonization method with multiple auxiliary variables. We find that in a simplified U(1) 
version of the model factorization holds, whereas in the full SU(3)-flavor version of the model factorization is broken by 
terms which are related to the 't Hooft interaction. 

\end{abstract}

\pacs{~ 12.38.-t, 12.38.Aw, 11.15.Tk, 12.39.Fe,12.39.Mk}

\maketitle

\section{Introduction}

One of the most interesting features of QCD is its non trivial vacuum, in which there is a complex configuration of quark and gluon fields.
The detailed understanding of these  vacuum fields will presumably be achieved  with lattice QCD calculations. However, so far we know only some 
expectation values of these fields, the condensates. In particular there are recent lattice calculations \cite{lepage,rede2}   
of the  quark condensate 
$\langle \bar{q} q \rangle$ and also phenomenological estimates of the quark  condensate and of the gluon condensate 
$\langle F_{\mu \nu}^a F^{a \mu \nu} \rangle$ \cite{nari1}. 
In  QCD sum rules (QCDSR) \cite{shif,na1,cola,na2,cola2,nnl} the mass of a hadron is extracted from the 
two-point correlation function, on  which we perform an operator product 
expansion (OPE). As a 
result of the OPE, in the end of the calculation the mass of the  hadron is written as a series of terms involving condensates of increasing order, 
starting with the quark condensate. Condensates of higher order give a decreasing contribution to the mass. 
In OPE diagrams for traditional hadrons, the number of quark lines is two and 
three for mesons and baryons respectively.  
Each of these lines, once we ``cut it'',  can give rise to a $\langle q \bar{q} \rangle$ condensate. When two lines are
simultaneously cut, we have the four-quark condensate $\langle q \bar{q}  q \bar{q} \rangle$.
Over the last ten years we have observed new multiquark ``exotic'' states, 
such as the $X(3872)$ \cite{nnl}. For these states the OPE diagrams contain more lines and higher order quark condensates may 
play a more important role \cite{msnn}. The four-quark condensate has not yet been  computed in lattice QCD 
calculations. Since it is a property of the vacuum, 
this, as well as other condensates, is a universal parameter,
 i.e., it is the same in every calculation of any hadron mass. 
 Because of this
 universality, we can 
calculate a certain number of hadronic masses and, fitting the QCDSR results to data, we can determine the values of the most relevant 
(and lowest order) condensates. 
There are large uncertainties in this procedure. In order to reduce the number of unknown parameters, the factorization hypothesis has 
often been 
used, according to which we can write:
\begin{equation}
\langle q \bar{q}  q \bar{q} \rangle = \, A \, \langle q \bar{q} \rangle^2
\label{fac}
\end{equation}
where $A$ is a constant. There is no justification of the above formula based on first principle calculations. Within some approximations,   
assuming the vacuum saturation hypothesis or resorting to the large Nc limit the four-quark condensates can be factorized into products of 
condensates with two quark operators \cite{kampfer}.  In \cite{zong} an attempt was made to directly calculate  the dimension four quark 
condensate and it was shown that, if nonperturbative effects are small, the factorization (\ref{fac}) is valid. In \cite{gomez}
the authors  addressed this issue within the low-energy representation of the condensates provided by Chiral Perturbation Theory. They obtained a 
formal, model-independent, proof of the non-validity of the factorization for the low-energy sector of QCD.

In the present work we address the issue  of factorization in the framework of an extension of the
 Nambu Jona-Lasinio (NJL) model, which includes higher order interactions.
 The inclusion of an eighth order interaction term
 leads to the appearance of the four-quark condensate besides the usual quark-antiquark condensate
by considering the auxiliary variable method.
This method is usually employed to investigate the emergence of the scalar chiral condensate
\cite{NJL,njl2,ralk} and it will be extended here to incorporate  higher order condensates. 
 This extension requires the 
introduction of new auxiliary variables. 
As it will be seen, the   gap equations mix the dimension two and dimension four 
condensates and establish a relation between them.

The advantage of using the NJL model is that it allows us not only to test the validity of (\ref{fac}) but also to identify the 
dynamical origin of the 
factorization breaking, which can be attributed to certain types of interactions.

\section{ The NJL model with U(1) symmetry}

Before starting  the full discussion of interacting quarks with three flavors, it is useful first to explore a simpler version of the NJL model
with one species of fermion. This U(1) prototype has many of the important features (such as spontaneous chiral symmetry breaking) which we 
encounter in the full flavor SU(3) version.

In its more traditional version \cite{NJL}, the Nambu Jona-Lasinio model describes the interaction between quarks in the low energy regime 
through pointlike four-fermion vertices. Its great merit is to incorporate chiral symmetry and  spontaneous chiral  symmetry breaking. It is 
also possible to include an 6th order term to account for 't Hooft interactions \cite{NJL}. We shall use here an extended 
version  \cite{osipov,osipov2,osipov3}  which includes also an 8th order interaction term. 
The U(1) Lagrangian density  including a   chiral invariant eighth order term is given by:
% \cite{osipov}:
\begin{equation} 
\mathcal{L} =  \bar{\psi} \left(i \slashed{\partial} - m_0 \right) \psi 
+ g_4 [(\bar{\psi} P_R \psi)(\bar{\psi} P_L \psi)]
+ g_8 [(\bar{\psi} P_R \psi)(\bar{\psi} P_L \psi)]^2
\label{lagrosip}
\end{equation}
where 
\begin{equation}
P_{R,L} = \frac{1}{2} (1 \pm \gamma_5)
\label{defproj}
\end{equation}
In the above Lagrangian, the second term is the usual NJL interaction and the third one is the  eighth order interaction term. 
It is interesting to notice that in the U(1) simplified model the instanton induced  't Hooft term introduces only a
correction to the current quark mass $m_0$ \cite{creutz}  and hence it can be absorbed into a 
redefinition of $m_0$, which in this model is a parameter anyway. Therefore we will neglect it in this section. 
As it will be seen in the next section, in the 
SU(3) flavor case, the   't Hooft interaction will introduce a sixth order term in the Lagrangian.

The generating functional for this simplified  model reads: 
\begin{eqnarray} 
\label{gfun}
Z =  \int {\cal D} [\bpsi, \psi] \, \exp  \; i \int d^4 x   \;  \left\{ 
\bpsi (x) \left( i \slashed{\partial} - m_0 \right) \psi 
+ \frac{g_4}{4} \left[ \left(\bpsi \psi \right)^2  -     \left( \bpsi  \gamma_5 \psi \right)^2 \right] 
 + \frac{g_8}{16} \left[ \left(\bpsi \psi \right)^2 -  \left( \bpsi  \gamma_5 \psi \right)^2  
                   \right]^2 
\right\} 
\end{eqnarray}
where a normalization constant is implicit. In what follows, we will apply  the bosonization prescription, 
introducing  scalar and pseudoscalar auxiliary fields. This is a generalization of the standard bosonization method. 
The path integral above is multiplied by a constant factor $N$ written as:
\begin{eqnarray}
N =  \int \; {\cal D} [\sigma] \; e^{- i \int d^4 x \frac{1}{2 \alpha}  \sigma^2(x)}  \;
     \int \; {\cal D} [\pi]    \; e^{- i \int d^4 x \frac{1}{2 \alpha}   \pi^2(x)}      \; 
     \int \; {\cal D} [\eta]   \; e^{- i \int d^4 x \frac{1}{2 \gamma}  \eta^2(x)}    \; 
     \int \; {\cal D} [\zeta]   \; e^{- i \int d^4 x \frac{1}{2 \delta} \zeta^2(x)}    \;
\label{mult} 
\end{eqnarray}
where  $\alpha$, $\gamma$ and $\delta$ are constants. 
We next perform unitary Jacobian shifts of the auxiliary variables such that the polynomial interactions cancel out.
For the Lagrangian density given in expression (\ref{lagrosip})
the appropriate shifts of the variables are given by:
\begin{equation}
\sigma^2  \rightarrow  \left( \sigma - a \, \bar{\psi} \psi  \right)^2
\label{shifsig}
\end{equation}
\begin{equation}
\pi^2  \rightarrow  \left( \pi - a \, \bar{\psi} \gamma_5 \psi  \right)^2
\label{shifpi}
\end{equation}
\begin{equation}
\eta^2  \rightarrow  \left( \eta - r  \,   (\bar{\psi} \psi)^2   \right)^2
\label{shifeta}
\end{equation}
\begin{equation}
\zeta^2  \rightarrow  \left( \zeta -   u \, (\bar{\psi} \psi)^2   - t \,   (\bar{\psi} \gamma_5 \psi)^2   \right)^2
\label{shifzeta}
\end{equation}
where $a$, $r$, $u$ and $t$ are constants. In Eq. (\ref{gfun}) chiral symmetry requires that the bilinears
$\bar{\psi} \psi $  and $\bar{\psi} \gamma_5 \psi$ appear with the same weigth. The auxiliary variables $\sigma$ 
and $\pi$ are also ``chiral partners'' and for this reason they appear multiplied by the same constant 
$1/2 \alpha$ in (\ref{mult}) and then, in (\ref{shifsig}) and (\ref{shifpi}) they are shifted by the mentioned 
bilinears multiplied by the same constant $a$.

After the multiplication by (\ref{mult}) and the shifts (\ref{shifsig}), (\ref{shifpi}), (\ref{shifeta}) and (\ref{shifzeta})
the generating functional (\ref{gfun}) can be rewritten as:
\begin{eqnarray}
Z &=& N  \int  {\cal D} [\sigma, \pi, \eta, \zeta]  \int {\cal D} [\bpsi, \psi] \, 
\exp  \; i \int d^4 x \Big[ \bpsi (x) \left(i \slashed{\partial} - m_0 + \frac{a \sigma}{\alpha} + \frac{ a \pi \gamma_5}{\alpha} \right) \psi + \nonumber \\ 
&&+ \left(  \frac{g_4}{4} - \frac{a^2}{2 \alpha} + \frac{r \eta}{\gamma} + \frac{u \zeta}{\delta} \right)  \left( \bpsi \psi \right)^2 +  
\left( - \frac{g_4}{4} - \frac{a^2}{2 \alpha} -\frac{t \zeta}{\delta} \right) \left( \bpsi \gamma_5 \psi \right)^2 
+ \left( \frac{g_8}{16} - \frac{r^2}{2 \gamma} -    \frac{u^2}{2 \delta} \right) \left( \bpsi \psi \right)^4  +  \nonumber \\
&&+ \left( \frac{g_8}{16} - \frac{t^2}{2 \delta}  \right)  \left( \bpsi \gamma_5 \psi \right)^4 + 
\left( \frac{g_8}{4} - \frac{u t}{2 \delta}  \right) \left( \bpsi \psi \right)^2 \left( \bpsi \gamma_5 \psi \right)^2 
 - \frac{\sigma^2}{2 \alpha}  - \frac{\pi^2}{2 \alpha}  
- \frac{\eta^2}{2 \gamma}  - \frac{\zeta^2}{2 \delta} \Big], 
\label{Z1}
\end{eqnarray}
As it can  easily be seen, the fermionic self  interaction terms   cancel out if the following relations are satisfied:
\begin{eqnarray} 
\label{cancel}
 \frac{g_4}{4} - \frac{a^2}{2 \alpha} + \frac{r \eta}{\gamma} + \frac{u \zeta}{\delta} = 0,
\\
 - \frac{g_4}{4} - \frac{a^2}{2 \alpha} -\frac{t \zeta}{\delta} = 0,
\\
 \frac{g_8}{16} - \frac{r^2}{2 \gamma} -    \frac{u^2}{2 \delta} = 0,
\\
\frac{g_8}{16} - \frac{t^2}{2 \delta} = 0,
\\
\frac{g_8}{4} - \frac{u t}{2 \delta} =  0,
\\
\end{eqnarray}
%The above system has five equations with five unknowns,  $a$, $b$, $r$, $u$ and $t$, which can be easily determined. 
Expression (\ref{Z1}) becomes then:
\begin{eqnarray} \label{Z-4-6-8}
Z &=& N  \int  {\cal D} [\sigma, \pi, \eta, \zeta]  \int {\cal D} [\bpsi, \psi] \, 
\exp  \; i \int d^4 x \Big[ \bpsi (x) \left(i \slashed{\partial} - m_0 + \frac{a \sigma}{\alpha} + 
\frac{ a \pi \gamma_5}{\alpha} \right) \psi  - \frac{\sigma^2}{2 \alpha}  - \frac{\pi^2}{2 \alpha}  
- \frac{\eta^2}{2 \gamma}  - \frac{\zeta^2}{2 \delta} \Big], 
\end{eqnarray}
The integration over the fermion field can be performed yielding:
\begin{eqnarray} \label{sigma-1}
Z = && N'    \int  {\cal D} [\sigma, \pi, \eta, \zeta]    \; exp  \Big[   i  \int d^4 x 
\left( - \frac{\sigma^2}{2 \alpha}  - \frac{\pi^2}{2 \alpha}  
- \frac{\eta^2}{2 \gamma}  - \frac{\zeta^2}{2 \delta}  \right)
+ i \, \mbox{Tr} \ln \left( i  \slashed{\partial} - m_0 + \frac{a \sigma}{\alpha} + \frac{ a \pi \gamma_5}{\alpha} \right) \Big]
,
\end{eqnarray}
where $N'$ differs from $N$ by a constant factor
and  Tr stands for traces of all indices: color, flavor and  Lorentz indices and a space-time integration. To arrive at this expression the 
following identity  was used: $\det A  = \exp \mbox{Tr} \ln A$. 
From the above expression we can read the effective action:
\begin{equation}
{\cal S}_{eff} =   i \, \mbox{Tr} \ln \left( i  \slashed{\partial} - m^* \right) 
- \int d^4 x \; \left( 
\frac{\sigma^2}{2 \alpha}  + \frac{\pi^2}{2 \alpha} + \frac{\eta^2}{2 \gamma}  + \frac{\zeta^2}{2 \delta}  
\right)
\label{leff}
\end{equation}
where the following effective mass was defined:
\begin{equation}
m^* =   m_0 - \frac{a \sigma_0}{\alpha} - \frac{ a \pi_0 \gamma_5}{\alpha} ,
\label{mstar}
\end{equation}
where the the subscripts $_0$ in the fields stand for their constant (mean-field) 
values in the vacuum.
The gap equations are found by the stationary conditions:
\begin{equation}
\frac{ \delta {\cal S}_{eff}}{\delta \sigma} \Big|_{\sigma= <\sigma>} = 0,  \;\;\;\;\;
\frac{ \delta {\cal S}_{eff}}{\delta \pi} \Big|_{\pi= <\pi>} = 0 , \;\;\;\;\;
\frac{ \delta {\cal S}_{eff}}{\delta \eta} \Big|_{\eta= <\eta>} = 0 , \;\;\;\;\;
\frac{ \delta {\cal S}_{eff}}{\delta \zeta} \Big|_{\zeta= <\zeta>} = 0, \;\;\;\;\;
\label{gaps}
\end{equation}
They turn out to be:
\begin{equation}
< \bar{\psi} \psi > =   i \,  \mbox{Tr} \left( \frac{1}{i \slashed{\partial} - m^*} \right) ,
\label{gapsig}
\end{equation}
\begin{equation}
< \bar{\psi} \gamma_5 \psi > = 0 ,
\label{gapi}
\end{equation}
\begin{equation}
< \bar{\psi}  \psi \bar{\psi}  \psi > =  < \bar{\psi} \psi >^2,
\label{gapeta}
\end{equation}
where we have used the  definitions $<\sigma> = a < \bar{\psi} \psi >$, $<\pi> = a < \bar{\psi} \gamma_5 \psi >$, 
$ <\eta> =  r < \bar{\psi}  \psi \bar{\psi}  \psi >$ and 
$ <\zeta> =  u < \bar{\psi}  \psi \bar{\psi}  \psi > \, + \, t < \bar{\psi} \gamma_5 \psi \, \bar{\psi} \gamma_5 \psi > $.  
The condensates appear as the values of the auxiliary fields in the stationary configurations. Equation (\ref{gapsig}) is the
gap equation obtained in \cite{NJL}, where the U(1) model has also been studied.  From the above exercise we conclude that 
the factorization hypothesis holds in the U(1) model,  where there is no 6th order interaction !

\section{The NJL model with SU(3) symmetry}

We now consider the SU(3) version of the model, as in Ref. \cite{osipov}. The  
Lagrangian is given by 
\begin{eqnarray}
\mathcal{L} = && \bar{\psi}_i \left(i \slashed{\partial} - m_0 \right) \psi_i  +
g_4
[(\bar{\psi}_i P_R \psi_m)(\bar{\psi}_m P_L \psi_i)]
+
\kappa (\det \bar{\psi}P_L \psi + \det \bar{\psi} P_R \psi)
+8
g_1 [(\bar{\psi}_i P_R \psi_m)(\bar{\psi}_m P_L \psi_i)]^2
\nonumber
\\
&&
+ 16g_2(\bar{\psi}_i P_R \psi_m)(\bar{\psi}_m P_L \psi_j)(\bar{\psi}_j P_R \psi_k)
(\bar{\psi}_k P_L \psi_i),
\label{lagrosip3}
\end{eqnarray}
where $i$, $j$,...etc, are flavor indices running from 1 to 3 and the summation over repeated indices is understood.
In the above expression  the  first  interaction term is the usual fourth order  NJL and the second 
is  the  't Hooft  determinantal term due to instantons. This term   reduces to a sixth order term 
in the $SU(3)$ case studied below. The last two pieces of (\ref{lagrosip3})  are the most general chiral invariant 
eighth order interaction terms, which may be attributed to instanton physics \cite{simonov,simonov2} or to 
vacuum polarization effects \cite{eNJL,enjl2}.
It is interesting to note that whereas the  't Hooft interaction makes the ground state of the model unstable
these eighth order interactions restore the stability of the vacuum \cite{osipov}. Using the definition of the 
projectors (\ref{defproj}), the above Lagrangian can be rewritten in a less compact but more useful form:
\begin{eqnarray} 
{\cal L} &=& \bar{\psi} \left(i \slashed{\partial} - m_0 \right) \psi 
+
 \frac{g_4 }{2}\left( \bpsi \lambda_b \psi)^2 + 
( \bpsi \lambda_c \gamma_5 \psi )^2 \right)
+ 
 \frac{\kappa}{48} d_{abc}  \bpsi \lambda_a \psi
\left(   \bpsi \lambda_b \psi  \bpsi \lambda_c \psi
- 3  \bpsi \lambda_b \gamma_5  \psi \bpsi \lambda_c \gamma_5 \psi 
\right)
\nonumber
\\
&& 
+ \frac{g_1}{2}
\left( ( \bpsi \lambda_a \psi )^2 +  (\bpsi \lambda_a \gamma_5 \psi )^2
\right)^2
\nonumber
\\
&& 
+ \frac{g_2}{8} d_{abe} d_{cde} 
\left[
 \bpsi \lambda_a \psi   \bpsi \lambda_b \psi  \bpsi \lambda_c \psi  \bpsi \lambda_d \psi
+ 
 \bpsi \lambda_a \gamma_5 \psi   \bpsi \lambda_b \gamma_5 \psi  \bpsi \lambda_c \gamma_5 \psi 
 \bpsi \lambda_d \gamma_5 \psi
+ 2 
 \bpsi \lambda_a \psi   \bpsi \lambda_b \psi  \bpsi \lambda_c \gamma_5 \psi  \bpsi \lambda_d \gamma_5 \psi
\right]
\nonumber
\\
&& + \frac{g_2}{2} f_{ace} f_{bde}
 \bpsi \lambda_a \psi   \bpsi \lambda_b \psi  \bpsi \lambda_c \gamma_5 \psi  \bpsi \lambda_d  \gamma_5\psi
\label{lagrafin}
\end{eqnarray}
where the indices $i,j,...$ etc of the SU(3) fundamental representation have been omitted and only the indices 
$a,b,...$ etc of the adjoint representation are shown. 
%The last term can be rewritten as:
%\begin{eqnarray}
% \frac{g_2}{2} f_{abe} f_{cle}
% \bpsi \lambda_a \psi   \bpsi \lambda_c \psi  \bpsi \lambda_b \gamma_5 \psi  \bpsi \lambda_l  \gamma_5\psi 
%&=&
%\frac{g_2}{2} \left[
%\frac{2}{3} \left( \delta_{ac} \delta_{bl} - \delta_{al} \delta_{bc} \right)
% \bpsi \lambda_a \psi   \bpsi \lambda_c \psi  \bpsi \lambda_b \gamma_5 \psi  \bpsi \lambda_l  \gamma_5\psi 
%\right.
%\nonumber
%\\
%&& \left. + \left(
%d_{ace} d_{ble} - d_{dle} d_{bce}
%\right) 
% \bpsi \lambda_a \psi   \bpsi \lambda_c \psi  \bpsi \lambda_b \gamma_5 \psi  \bpsi \lambda_l  \gamma_5\psi 
%\right]
%\end{eqnarray}

In what follows we will repeat the steps  of the previous section but  now  new  auxiliary variables 
will be needed to account for the sixth and   two different eighth order interactions. 
The generating functional now reads: 
\begin{eqnarray} 
\label{gfun2}
Z =  \int {\cal D} [\bpsi, \psi] \, \exp  \; i \int d^4 x    \; \mathcal{L} 
\end{eqnarray}
with $\mathcal{L}$ given by (\ref{lagrafin}). 
In the  above  Lagrangian  there are terms with eight fields which will give rise to four-quark condensates. Since we want to 
make sure that we are not using any factorization hypothesis from the beginning, we must introduce the following four auxiliary variables: 
$\bar{\psi} \lambda_a \psi$, $\bar{\psi} \lambda_a \gamma_5 \psi $, $\bar{\psi} \lambda_a \psi \bar{\psi} \lambda_a \psi$ and 
$\bar{\psi} \lambda_a \gamma_5 \psi \bar{\psi} \lambda_a \gamma_5 \psi $. 
%$\bar{\psi} \lambda_a  \psi \bar{\psi} \lambda_a \gamma_5 \psi $. 
For the sake of 
convenience, the last two variables will be recombined and will become two new variables. The final four auxiliary variables will then 
be $\varphi_a$, $\pi_a$,   $\phi_a$ and $\omega_a$ and we will write four gap equations for them. 
We will not be able to recover the same results found in \cite{ralk} because here we include terms with eight
fields whereas in \cite{ralk} there are no such terms. Moreover the four quark condensate is for us an independent dynamical variable.
The bosonization method used here relies on cancellations which come from  shifts in  the Gaussian integrals of the auxiliary variables.
Given that we need at least four auxiliary variables to describe the four condensates considered
(two 2-quark and two 4-quark condensates), we could proceed by shifting these four auxiliary variables by the appropriate number
of quark bilinears and quadrilinears in such a way  that all the Lagrangian interactions would be cancelled out.
However, it turns out that this is not possible because crossed terms from the shifts  introduce  spurious Lagrangian
interactions which were not present in the original model. Therefore, further auxiliary variables must be introduced so that the
individual shifts for the auxiliary variables yield the  expected terms for the cancellations.
In the end all but four can be eliminated, resulting in the four gap equations.

Following the steps of the preceding section we multiply the above expression by a constant factor 
written as:
\begin{eqnarray} 
1 &=& N' 
% \int D \; [\sigma_a] \; e^{ i \int d^4 x  [\sigma_a (\phi_a - \xi_a) 
% + \theta_a (\phi_a - \omega_a) ] 
% }
%\int D [
%\tilde{\theta}_a,\tilde{\sigma}_a] \;  e^{ i \int d^4 x [
% \tilde{\theta}_a (\omega_a - \Omega_a) 
%+ \tilde{\sigma}_a (\xi_a - \rho_a)]  }
%\nonumber
%\\
%&& \times
\int \; {\cal D} [\phi_a] \; e^{- i \int d^4 x \frac{1}{2 \alpha} \phi_a^2   }  \;
\int \; {\cal D} [\varphi_a, \pi_a] \; e^{- i \int d^4 x \frac{1}{2 \beta} (\varphi^2_a + \pi_a^2)} 
\int \; {\cal D} [\omega_a] \; e^{- i \int d^4 x \frac{1}{2 \delta }  \omega_a^2 } 
%\nonumber
%\\
%&&
% \times
%\int \; {\cal D} [\xi_a] \; e^{- i \int d^4 x \frac{1}{2  \gamma} \xi^2_a } \;
% \int \; {\cal D} [\chi_a] \;   e^{- i \int d^4 x \frac{1}{2 d_2}   \chi_a^2 }  \;
%  \;
%% \int \; {\cal D} [\Omega] \; e^{- i \int d^4 x \frac{1}{2 \alpha}  \Omega^2 }
%\nonumber
%\\
%&& 
%\int \; {\cal D} [\Omega] \; e^{- i \int d^4 x \frac{1}{2 \alpha}  \Omega^2 }  \; .
%\int \; {\cal D} [\rho_a] \; e^{- i \int d^4 x \frac{1}{2 c_3 }  \rho_a^2  }  .
\label{constfac}
\end{eqnarray}
In this expression there are four auxiliary variables $\varphi_a$, $\pi_a$, $\phi_a$,  and   $\omega_a$.
We next rewrite the action introducing  shifts analogous to those of expressions
(\ref{shifsig}), (\ref{shifpi}), (\ref{shifeta}) and (\ref{shifzeta}). Following the steps of the last 
section, we would now be tempted to introduce the same shifts (except for the flavor indices) and they would lead 
to the shifted variables $\varphi_a = \bar{\psi} \lambda_a \psi $,  $\pi_a = \bar{\psi} \lambda_a \gamma_5 \psi $, 
$\phi = \bar{\psi} \lambda_a \psi \bar{\psi} \lambda_a \psi $ and  
$\omega = \bar{\psi} \lambda_a \gamma_5 \psi \bar{\psi} \lambda_a \gamma_5 \psi $.  After the extremization of the 
action, the obtained values for these variables would be the condensates. We might then expect to find relations among 
them such as, for example, $\phi \propto \varphi_a^2$, etc... However, as it is can be easily checked, with these simple 
shifts the needed cancellation of the higher power fermion bilinears in the action will not happen! This is mainly due to 
the appearance of the SU(3) structure constants  in (\ref{lagrafin}).  We will develop a strategy to circumvent this problem. 
We start noting that:
\begin{equation} 
\phi_a^2 \, = \, \frac{3}{5 } ( d_{abc} \phi_a)( d_{dbc} \phi_d)   
\end{equation} 
and also that:
\begin{equation}
\omega_a^2 \, = \, \frac{1}{  N_f } ( f_{abc} \omega_a)( f_{dbc} \omega_d)  .
\end{equation}
We now introduce the shifts which will lead  to a cancellation of both the 't Hooft  and  the eighth order  
interaction terms (with $g_1$ and $g_2$).  They are: 
\begin{eqnarray} 
\label{shifts-gen}
- \frac{1}{2 \beta} (\varphi^2_a + \pi_a^2)  \to   && -  \frac{1}{2 \beta} 
\left( (\varphi_a - \epsilon_s (\bpsi \lambda_a \psi) )^2 + (\pi_a - \epsilon_p (\bpsi i \gamma_5 \lambda_a \psi) )^2
\right) ,
\\
\label{phi}
- \frac{1}{2 \alpha} \phi_a^2 = -  \frac{3}{10 \alpha } ( d_{abc} \phi_a)( d_{dbc} \phi_d)   
\to  && - \frac{3}{10 \alpha  } \left(
\phi_a d_{abc} - T_a d_{abc} (\bpsi \lambda_e \psi)^2
- \epsilon_1 f_{bco} f_{ode} \bpsi \lambda_d \psi \bpsi \lambda_e \psi
 - \epsilon_3 d_{abc} (\bpsi \lambda_{a} \psi)
\right.
\nonumber
\\
&& 
\left.  - \epsilon_2 \delta_{bc}  (\bpsi \lambda_e \psi)^2
\right)^2 ,
\\
\label{shifomega}
- \frac{1}{2  \delta}  \omega_a^2  = -  \frac{1}{ 2 \delta  N_f } ( f_{abc} \omega_a)( f_{dbc} \omega_d)     
\to && - \frac{1}{2 \delta N_f} \left( \omega_a f_{abc} - \epsilon_1 \sqrt{\frac{1}{32 \alpha \delta} } d_{abc} d_{ade} 
(\bpsi \lambda_d \psi)( \bpsi \lambda_e \psi ) - 
\epsilon_4 d_{abc} d_{ade} ( \bpsi \lambda_d \gamma_5 \psi) (\bpsi \lambda_e \gamma_5 \psi) 
\right.
\nonumber
\\
&& \left. 
- 
T_a f_{abc} (\bpsi \lambda_d \psi )^2 
- T_a  f_{abc} 
(\bpsi \lambda_d i \gamma_5 \psi)^2
-  \epsilon_5 d_{abc} ( \bpsi \lambda_a \psi)
\right)^2 ,
\label{omegaa}
\end{eqnarray}
where $\epsilon_s$, $\epsilon_p$, $\epsilon_1$, $\epsilon_2$, $\epsilon_3$,  $\epsilon_4$ and  $\epsilon_5$ 
are constants and $T_a$ is a  constant flavor vector. From (\ref{omegaa}) we have:
\begin{equation}
\omega_a f_{abc} =   \epsilon_1 \sqrt{\frac{1}{32 \alpha \delta} } d_{abc} d_{ade}  (\bpsi \lambda_d \psi)( \bpsi \lambda_e \psi ) 
                 + \epsilon_4 d_{abc} d_{ade} ( \bpsi \lambda_d \gamma_5 \psi) (\bpsi \lambda_e \gamma_5 \psi)  
                 + T_a f_{abc} \left[ (\bpsi \lambda_d \psi )^2  + (\bpsi \lambda_d i \gamma_5 \psi)^2 \right]
                 +  \epsilon_5 d_{abc} ( \bpsi \lambda_a \psi) 
\label{defomega}
\end{equation}
Multiplying the above equation by $f_{ebc}$ we find:
\begin{equation} 
\omega_e  =   T_e \left[ (\bpsi \lambda_d \psi )^2  +    (\bpsi \lambda_d i \gamma_5 \psi)^2 \right]
\label{omegamult} 
\end{equation} 
where we have used the following SU(3) algebra relations: 
$d_{abc}d_{ebc}= \frac{5}{3} \delta_{ae}$,  $f_{abc} f_{ebc} = N_f \delta_{ae}$, $\delta_{aa}=(N_f^2-1)$ and 
also that  $\delta_{bc} f_{abc}=\delta_{bc} d_{abc} = f_{abc}d_{dbc}=0$ . 
Multiplying the above expression by $T_e$ and dividing by $T.T$ we find:
\begin{equation}  
\omega =\frac{T_e \omega_e}{T.T}  =   (\bpsi \lambda_d \psi )^2  +    (\bpsi \lambda_d i \gamma_5 \psi)^2 
\label{omegapuro}  
\end{equation}
The variable $\omega$ is the normalized average of $\omega_e$ over the SU(3) directions and is directly related to 
the combination of four quark condensates   and it is the SU(3) analogue of the $\zeta$ variable defined in the last 
section (\ref{shifzeta}).  For completeness we derive the equivalent definition of $\phi$. From (\ref{phi})
we have:
\begin{equation}
\phi_a d_{abc} = T_a d_{abc} (\bpsi \lambda_e \psi)^2 
                 + \epsilon_1 f_{bco} f_{ode} \bpsi \lambda_d \psi \bpsi \lambda_e \psi  + \epsilon_3 d_{abc} (\bpsi \lambda_{a} \psi) 
+ \epsilon_2 \delta_{bc}  (\bpsi \lambda_e \psi)^2
\label{defphi}
\end{equation}
Multiplying this equation by $d_{ebc}$ we have:
\begin{equation} 
\frac{5}{3} \phi_e  = \frac{5}{3} T_e  (\bpsi \lambda_e \psi)^2  +  \frac{5}{3} \epsilon_3 (\bpsi \lambda_{e} \psi)
\label{defphi} 
\end{equation} 
Multiplying this expression by $T_e$ and dividing by $T.T$ we have:
\begin{equation}  
\phi =  \frac{T_e \phi_e}{T . T}  =   (\bpsi \lambda_e \psi)^2   +  \frac{T_e (\bpsi \lambda_{e} \psi) \epsilon_3 }{T . T } 
\label{phipuro}  
\end{equation} 
with $\phi$ being the analogue of  $\eta$ defined above in (\ref{shifzeta}).   
The shifts above however do not cancel out all the sixth and eighth order terms.
Now we perform the following substitution in the first term of (\ref{constfac}), quadratic in $\phi_a$
\begin{equation} 
e^{- i \int d^4 x \frac{1}{2 \alpha} \phi_a^2}  = e^{- i \int d^4 x \frac{1}{2 \mu} \phi_a^2}
 \int D [\xi_a]  e^{- i \int d^4 x \,  \frac{\xi_a^2}{2 \gamma} } 
\int D [\sigma_a] e^{i \int d^4 x \, \sigma_a (\phi_a - \xi_a) },
\label{deltanew}
\end{equation}
where the second integral is the delta function $\delta [ \phi_a - \xi_a ]$ and $1/\alpha = 1/\mu + 1/\gamma$. 
We have introduced the new auxiliary variable $\xi_a$  which will be integrated out later.
The shift that allows for the cancellation of the remaining interactions is the folllowing:
\begin{eqnarray} \label{shift-2nd}
\frac{1}{2 \gamma} \xi_a^2 \, = \, \frac{1}{ 2 \gamma N_f} ( f_{abc} \xi_a)( f_{dbc} \xi_d) 
\to  &&
\frac{1}{ 2  \gamma N_f}
 \left( f_{abc} \xi_a -  d_{abc} d_{ade} \epsilon_6 (\bpsi \lambda_d \psi) (\bpsi \lambda_e \psi) 
+ f_{abc} T_a   (\bpsi \lambda_{a'} \psi)^2
\right.
\nonumber
\\
&& \left. 
-  \epsilon_3 d_{a'bc}   (\bpsi \lambda_{a'} \psi) - \epsilon_3 f_{a'bc}   (\bpsi \lambda_{a'} \psi)
\right)
\nonumber
\\
&&
\times
\left( f_{dbc} \xi_d -  d_{dbc} d_{def}  \epsilon_6 (\bpsi \lambda_e \psi) (\bpsi \lambda_f \psi) 
+  f_{abc} T_a  (\bpsi \lambda_{a'} \psi)^2
\right.
\nonumber
\\
&& \left.
-  \epsilon_3 d_{a'bc}   (\bpsi \lambda_{a'} \psi) - \epsilon_3 f_{a'bc}   (\bpsi \lambda_{a'} \psi)
\right) ,
\end{eqnarray}
%For the variable $\xi_a$  the corresponding shift is given by:
%$\xi_d \to \xi_d  - \epsilon_3  (\bpsi \lambda_{a'} \psi)$.
The shift above introduces a shift in the term $\sigma_a (\phi_a - \xi_a)$ which can be carried out remembering that:
%\to \sigma_a (\phi_a - \xi_a - \epsilon_3  (\bpsi \lambda_{a'} \psi))$.
\begin{equation}
\sigma_a (\phi_a - \xi_a) = \frac{f_{abc} \sigma_a}{N_f} ( f_{dbc} \phi_d - f_{dbc} \xi_d)
\end{equation}
%After this last shift we can perform the integration on $\sigma_a$ and then on $\xi_a$. 
With the variables and the shifts defined above our strategy will be to  compute  the cancellations of the
higher powers of fermion fields and the integration over these  fields. Then, extremizing the 
action with respect to the four relevant auxiliary variables, we will obtain gap equations for 
$\varphi_a$, $\pi_a$, $\phi_a$ and $\omega_a$ and solve them. The expressions for $\phi_a$ and $\omega_a$ 
will then be converted into expressions for $\phi$ and $\omega$ simply by being multiplied by 
 $T_a/T.T$. Finally, solving the gap equations we will find relations between the 
different condensates. 

Inserting (\ref{lagrafin}), (\ref{constfac}) and  (\ref{shift-2nd}) into (\ref{gfun2})  we find:
\begin{eqnarray}
Z &=&  \int {\cal D} [\bpsi, \psi]  \;  \int {\cal D} [\sigma_a,\phi_a,\varphi_a,\pi_a,\xi_a,\omega_a]
\exp \; i \int d^4 x \;
\left[ 
\bpsi \left( 
 i \slashed{\partial} - \left( m_0 - 
\frac{\epsilon_s  \varphi_a \lambda_a}{\beta} - \frac{\epsilon_p i \gamma_5 \lambda_a \pi_a}{\beta}
- \frac{\epsilon_3  \xi_a \lambda_a}{ \gamma} - \frac{\epsilon_3  \phi_a \lambda_a}{ \alpha}  
 \right) \right) \psi
\right.
\nonumber
\\
&& 
\left.
%+ \sigma_a (\phi_a - \xi_a - \epsilon_3  (\bpsi \lambda_{a'} \psi))
+ \frac{f_{abc} \sigma_a}{N_f} ( f_{dbc} \phi_d -  
\left( f_{dbc} \xi_d -  d_{dbc} d_{def}  \epsilon_6 (\bpsi \lambda_e \psi) (\bpsi \lambda_f \psi) 
+  f_{abc} T_a  (\bpsi \lambda_{a'} \psi)^2 -  \epsilon_3 d_{a'bc}   (\bpsi \lambda_{a'} \psi) - \epsilon_3 f_{a'bc}   
(\bpsi \lambda_{a'} \psi) \right)
)
\right.
\\
&&
\left. 
- \frac{\varphi_a^2 + \pi_a^2}{2 \beta} - \frac{\phi_a^2}{2 \mu} - \frac{\omega_a^2}{2 \delta}
-\frac{\xi_a^2}{2 \gamma}
-  \left(  \frac{1 }{\alpha} \epsilon_3 - \frac{1}{\gamma}
\epsilon_3 \right) T_a (\bpsi \lambda_{a'} \psi) ( (\bpsi \lambda_{b} \psi) )^2
\right.
\nonumber
\\
&& 
\left.
-  \left(  \frac{T_a^2}{\delta} - g_1  \right)
\left( \bpsi \lambda_a \gamma_5 \psi \right)^4
- 
\left( \frac{T_a^2}{\delta} -  g_1  \right) \left( \bpsi \lambda_a  \psi \right)^2 
\left( \bpsi \lambda_b \gamma_5 \psi \right)^2
-
\left( 
\frac{T_a^2}{2 \delta} + \frac{T_a^2}{2 \alpha} + \frac{T_a^2}{2 \gamma} 
 - \frac{(N_f^2-1) }{2 N_f \alpha} \epsilon_2^2
 - 
\frac{g_1}{2} \right) \left( \bpsi \lambda_a  \psi \right)^4 
\right.
\nonumber
\\
&& 
\left.
-
\left(   \frac{\epsilon_p^2 }{2 \beta}  - 
g_4  -   \frac{ T_a \cdot \omega_a}{ \delta} \right)
\left(   \bpsi \lambda_b  \gamma_5  \psi \right)^2
- \left(  \frac{\epsilon_6^2 5/3}{2 N_f\gamma} + \frac{ \epsilon_1^2 5/3}{2 N_f \delta} 
\frac{1}{32 \alpha \delta}  - \frac{ g_2}{8}
\right) d_{abc} d_{aef} \left(   \bpsi \lambda_b  \psi \right) \left(   \bpsi \lambda_c \psi \right)
\left(   \bpsi \lambda_e  \psi \right) \left(   \bpsi \lambda_f \psi \right)
\right.
\nonumber
\\   
&&
\left. 
-  \left( \frac{\epsilon_4^2 5/3}{2 N_f \delta} -
\frac{g_2}{8} \right) d_{abc} d_{aef}
\left(   \bpsi \lambda_b \gamma_5  \psi \right) \left(   \bpsi \lambda_c  \gamma_5 \psi \right)
\left(   \bpsi \lambda_e  \gamma_5 \psi \right) \left(   \bpsi \lambda_f  \gamma_5 \psi \right)
\right.
\nonumber
\\    
&& 
\left. 
-
\left( 
\frac{\epsilon_4 \epsilon_1 5/3}{N_f \delta} 
\sqrt{\frac{1}{32 \alpha \delta}  } - \frac{2 g_2}{8} \right)
d_{abc} d_{aef}
\left(   \bpsi \lambda_b \gamma_5  \psi \right) \left(   \bpsi \lambda_c  \gamma_5 \psi \right)
\left(   \bpsi \lambda_e   \psi \right) \left(   \bpsi \lambda_f  \psi \right)
\right.
\nonumber
\\   
&& 
\left.
-  \left( \frac{3 N_f \epsilon_1^2}{10 \alpha }  -  g_2  \right)
f_{abc} f_{aef}
\left(   \bpsi \lambda_b  \psi \right) \left(   \bpsi \lambda_c  \gamma_5 \psi \right)
\left(   \bpsi \lambda_e   \psi \right) \left(   \bpsi \lambda_f  \gamma_5 \psi \right)
\right.
\nonumber
\\    
&& 
\left. 
 - \left( 
\frac{5/3 \epsilon_6 \epsilon_3}{\gamma N_f}  + \frac{\epsilon_4   \epsilon_5  5/3}{\delta N_f}
-  \frac{ \kappa}{48} \right) 
d_{bce} 
\left(   \bpsi \lambda_b   \psi \right) \left(   \bpsi \lambda_c   \psi \right)
\left(   \bpsi \lambda_e   \psi \right)  
\right.
\nonumber
\\      
&& 
\left. 
+
\left( 
 \frac{\epsilon_4 \epsilon_5  5/3}{\delta N_f}  +  \frac{3 \kappa}{48} \right)
d_{bce} 
\left(   \bpsi \lambda_b   \psi \right) \left(   \bpsi \lambda_c  \gamma_5  \psi \right)
\left(   \bpsi \lambda_e  \gamma_5  \psi \right)  
\right.
\nonumber
\\   
&& 
\left. 
+
\left( - \frac{\epsilon_s^2}{2 \beta} + 
 g_4 + \frac{\phi_a T_a}{  \alpha}  - \frac{\xi_a T_a}{  \gamma} 
-  \frac{ \epsilon_3^2 5/3 }{2 N_f \gamma} -
 \frac{\epsilon_3^2}{2\gamma}   - \frac{\epsilon_3^2 }{2 \alpha}
- 
\frac{\epsilon_5^2 5/3}{2 \delta} +  \frac{\omega_a T_a}{\delta}
\right)
\left(   \bpsi \lambda_b   \psi \right)^2
\right].
\label{zefin}
\end{eqnarray}
Next the variable $\sigma_a$ can be integrated out, yielding a delta function which implies that
$\xi_a =  \phi_a  + \epsilon_3  (\bpsi \lambda_{a} \psi)  + T_a (\bpsi \lambda_b \psi)^2 $.
The above expression becomes:
\begin{eqnarray}
Z &=&  \int {\cal D} [\bpsi, \psi]  \;  \int {\cal D} [\phi_a,\varphi_a,\pi_a,\omega_a]
\exp \; i \int d^4 x \;
\left[ 
\bpsi \left( 
 i \slashed{\partial} -\left( m_0 - 
\frac{\epsilon_s  \varphi_a \lambda_a}{\beta} - \frac{\epsilon_p i \gamma_5 \lambda_a \pi_a}{\beta}
- \frac{\epsilon_3  \phi_a \lambda_a}{ \gamma} - \frac{\epsilon_3  \phi_a \lambda_a}{ \alpha}  
 \right) \right)
\right.
\nonumber
\\
&& 
\left.
- \frac{\varphi_a^2 + \pi_a^2}{2 \beta} - \frac{\phi_a^2}{2 \alpha} - \frac{\omega_a^2}{2 \delta}
-  \left(  \frac{1 }{\alpha} \epsilon_3 - \frac{2}{\gamma}
\epsilon_3  \right) T_a (\bpsi \lambda_{a'} \psi) ( (\bpsi \lambda_{b} \psi) )^2
\right.
\nonumber
\\
&& 
\left.
-  \frac{1}{2} \left(  \frac{T_a^2}{\delta} - g_1  \right)
\left( \bpsi \lambda_a \gamma_5 \psi \right)^4
- 
\left( \frac{T_a^2}{\delta} -  g_1  \right) \left( \bpsi \lambda_a  \psi \right)^2 
\left( \bpsi \lambda_b \gamma_5 \psi \right)^2
-
\left( 
\frac{T_a^2}{2 \delta} + \frac{4 T_a^2}{ \alpha} + \frac{T_a^2}{2 \gamma}  
 - \frac{(N_f^2-1) }{2 N_f \alpha} \epsilon_2^2- 
\frac{g_1}{2} \right) \left( \bpsi \lambda_a  \psi \right)^4 
\right.
\nonumber
\\
&& 
\left.
-
\left(   \frac{ \epsilon_p^2 }{2 \beta}  - 
g_4  -   \frac{ T_a \cdot \omega_a}{ \delta} \right)
\left(   \bpsi \lambda_b  \gamma_5  \psi \right)^2
- \left(  \frac{\epsilon_6^2}{2 \gamma} + \frac{ \epsilon_1^2}{2 \delta}\frac{1}{32 \alpha \delta} 
 - \frac{3 N_f g_2}{20}
\right) d_{abc} d_{aef} \left(   \bpsi \lambda_b  \psi \right) \left(   \bpsi \lambda_c \psi \right)
\left(   \bpsi \lambda_e  \psi \right) \left(   \bpsi \lambda_f \psi \right)
\right.
\nonumber
\\   
&&
\left. 
-  \left( \frac{\epsilon_4^2 5/3}{2 N_f \delta} -
\frac{g_2}{8} \right) d_{abc} d_{aef}
\left(   \bpsi \lambda_b \gamma_5  \psi \right) \left(   \bpsi \lambda_c  \gamma_5 \psi \right)
\left(   \bpsi \lambda_e  \gamma_5 \psi \right) \left(   \bpsi \lambda_f  \gamma_5 \psi \right)
\right.
\nonumber
\\    
&& 
\left. 
-
\left( 
\frac{\epsilon_4 \epsilon_1 5/3}{N_f \delta} \sqrt{\frac{1}{32 \alpha \delta} }  - \frac{2 g_2}{8} \right)
d_{abc} d_{aef}
\left(   \bpsi \lambda_b \gamma_5  \psi \right) \left(   \bpsi \lambda_c  \gamma_5 \psi \right)
\left(   \bpsi \lambda_e   \psi \right) \left(   \bpsi \lambda_f  \psi \right)
\right.
\nonumber
\\   
&& 
\left.
-  \left( \frac{3 N_f \epsilon_1^2}{10 \alpha }  -  g_2  \right)
f_{abc} f_{aef}
\left(   \bpsi \lambda_b  \psi \right) \left(   \bpsi \lambda_c  \gamma_5 \psi \right)
\left(   \bpsi \lambda_e   \psi \right) \left(   \bpsi \lambda_f  \gamma_5 \psi \right)
\right.
\nonumber
\\    
&& 
\left. 
 - \left( 
\frac{5/3 \epsilon_6 \epsilon_3}{N_f \gamma}  + \frac{\epsilon_4   \epsilon_5  5/3}{\delta N_f}
-  \frac{ \kappa}{48} \right) 
d_{bce} 
\left(   \bpsi \lambda_b   \psi \right) \left(   \bpsi \lambda_c   \psi \right)
\left(   \bpsi \lambda_e   \psi \right)  
\right.
\nonumber
\\      
&& 
\left. 
+
\left( 
 \frac{\epsilon_4 \epsilon_5  5/3}{\delta N_f}  +  \frac{3 \kappa}{48} \right)
d_{bce} 
\left(   \bpsi \lambda_b   \psi \right) \left(   \bpsi \lambda_c  \gamma_5  \psi \right)
\left(   \bpsi \lambda_e  \gamma_5  \psi \right)  
\right.
\nonumber
\\   
&& 
\left. 
+
\left( - \frac{\epsilon_s^2}{2 \beta} + 
 g_4 + \frac{\phi_a T_a}{  \alpha}  -  \frac{\phi_a T_a}{  \gamma}
-  \frac{ \epsilon_3^2 5/3 }{2 N_f \gamma} -
 \frac{\epsilon_3^2}{2\gamma} 
-  \frac{\epsilon_3^2}{\gamma} 
- \frac{\epsilon_3^2}{2 \gamma}
- 
\frac{\epsilon_5^2 5/3}{2 \delta} +  \frac{\omega_a T_a}{\delta} 
 + \frac{\epsilon_3^2}{\gamma}  + \frac{\epsilon_3^2}{2 \gamma}
\right)
\left(   \bpsi \lambda_b   \psi \right)^2
\right].
\label{zefin}
\end{eqnarray}

From the above equation we can see that the desired cancellations of the self-interaction terms will happen  
if the following relations are satisfied:
\begin{eqnarray}
\;\;\;  2 \alpha &=& \gamma 
\\
\label{A-2}
\;\;\;\frac{T_a^2}{\delta} &=&  g_1 
\\ \label{A-3}
\;\;\;   \frac{T_a^2}{\alpha} 
%+ \frac{ \epsilon_6^2 (N_f^2-1)}{\gamma N_f}
 &=&  \frac{2 (N_f^2-1) }{17 N_f \alpha} \epsilon_2^2
\\      \label{B-1}
\;\;\;  \frac{\epsilon_6^2}{\gamma} + \frac{\epsilon_1^2}{\delta} \frac{1}{32 \alpha \delta}  &=& \frac{3 N_f}{20} g_2
\\      \label{B-2}
\;\;\;  \frac{\epsilon_4^2}{\delta} &=& \frac{6 N_f}{40} g_2
\\      \label{C-1} 
\;\;\;  \frac{\epsilon_1^2}{\alpha} &=& \frac{5}{3} g_2
\\      \label{D-1}
\;\;\;  \frac{\epsilon_6 \epsilon_3}{\gamma}  + \frac{\epsilon_4 \epsilon_5}{\delta} &=& \frac{ N_f}{80} \kappa
\\      \label{D-2}
\;\;\;  \frac{\epsilon_4 \epsilon_5}{\delta}  &=& -  \frac{3 N_f}{80} \kappa
\\      \label{E-2} 
\;\;\;  
\frac{ \epsilon_s^2 }{2 \beta} 
+  \frac{ \epsilon_3^2 5/3 }{2 N_f \gamma} +
 \frac{ \epsilon_3^2}{2 \gamma}
+ \frac{5/3 \epsilon_5^2}{2 \delta}   &=& \frac{T_a \cdot \phi_a}{  \gamma} + g_4  +  \frac{ T_a \cdot \omega_a}{ \delta}
\\      \label{E-1}
\;\;\;  \frac{\epsilon_p^2 }{\beta} &=& g_4  +  \frac{ T_a \cdot \omega_a}{ \delta}.
\end{eqnarray}
From the first identity, one  also has: $\mu = \gamma$.
The  last expression implies that  the shift parameters  $\epsilon_s$ and $\epsilon_p$ are 
actually functions of the auxiliary variables  $\phi_a$ and $\omega_a$. This is not a problem  
because the corresponding shifts for the fields $\varphi_a$ and $\pi_a$ 
in expression (\ref{shifts-gen}) still have unitary Jacobians. 
From expressions  (\ref{B-1}), (\ref{B-2}), (\ref{D-1}) and  (\ref{D-2})  there is a non-ambiguous identification of 
$\epsilon_5$ (and $\epsilon_3$) with the 't Hooft coupling constant $\kappa$. This will be a relevant issue. The generating 
functional of the  resulting effective theory becomes:
\begin{eqnarray} \label{Z-4-6-8}
Z &=& 
N' \int {\cal D}  [\bpsi, \psi] \;  \int {\cal D} [\phi_a,\varphi_a,\pi_a,\omega_a] \;   
%% \int {\cal D} \; 
\exp \;   i  \int d^4 x \left[  \bpsi \,  [ \, \slashed{\partial} 
-   \left( m_0 - 
\frac{\epsilon_s  \varphi_a \lambda_a}{\beta} - \frac{\epsilon_p i \gamma_5 \lambda_a \pi_a}{\beta}
- \frac{3 \epsilon_3  \phi_a \lambda_a}{ 2 \alpha}  
 \right) \, ] \, \psi 
\right.
\nonumber
\\
&&  \left.
- \frac{1}{2 \alpha} \phi_a^2 
- \frac{1}{2  \beta} ( \varphi_a^2  + \pi_a^2)   
- \frac{1}{2 \delta} \omega_a^2 
 \right] .
\end{eqnarray}
The integration over the quark fields yields  :
\begin{eqnarray} \label{sigma-2}
Z &=& N'' \int  {\cal D} [\phi_a, \varphi_a, \pi_a,\omega_a] \; exp  \left[  - i  \int d^4 x  
\left( \frac{1}{2 \mu}   \phi_a^2 
+
\frac{1}{\delta} \omega_a^2
+ \frac{1}{2 \beta}   (\varphi_a^2  +\pi_a^2)
\right) 
 \right] 
\nonumber
\\
&& exp  \left[    \mbox{Tr}  \ln \left( i \slashed{\partial} - 
\left( m_0 - 
\frac{\epsilon_s  \varphi_a \lambda_a}{\beta} - \frac{\epsilon_p i \gamma_5 \lambda_a \pi_a}{\beta}
- \frac{3 \epsilon_3  \phi_a \lambda_a}{ 2 \alpha} 
%% - \frac{\epsilon_3  \phi_a \lambda_a}{ \gamma}  
  \right)
\right)  \right] .
\end{eqnarray}
where the big  trace (Tr) is, as before, is  taken over all discrete indices of color, flavor and  Dirac,   and also implies a 
space-time integration.  From the above expressions we can read the following effective action:
\begin{eqnarray}
{\cal S}_{eff} &=&   \, - i  \mbox{Tr} \ln \left( i  \slashed{\partial} 
-  
\left( m_0 - 
\frac{\epsilon_s  \varphi_a \lambda_a}{\beta} - \frac{\epsilon_p i \gamma_5 \lambda_a \pi_a}{\beta}
-3  \frac{\epsilon_3  \phi_a \lambda_a}{2  \alpha} 
 \right)  \right) 
\nonumber
\\
&& - 
\int d^4 x \; \left(
 \frac{\phi^2_a}{2 \alpha} 
 + \frac{\omega^2_a}{2 \delta}
 +
 \frac{\varphi^2_a}{2 \beta}  + \frac{\pi^2_a}{2 \beta}  
\right).
\label{leff2}
\end{eqnarray} 
The gap equations for the variables $\phi_a, \omega_a,\pi_a$ and $ \varphi_a$ are the following: 
\begin{equation} 
\label{gap1}
\frac{\partial {\cal S}_{eff}}{\partial \varphi_a}  \Big|_{\varphi_a= <\varphi_a>} = 0   \,  \to \,   
<\varphi_a>  = -  \left[ \mbox{Tr'}  \frac{i}{ i \slashed{\partial} - { m^*}} \right] \,\, \lambda_a \epsilon_s ,
\end{equation}
\begin{equation}
\label{gap2}
\frac{\partial {\cal S}_{eff}}{\partial \pi_a}   \Big|_{\pi_a= <\pi_a>}  = 0  \, \to \,   
<\pi_a> =  - \left[ \mbox{Tr'}  \frac{i}{ i \slashed{\partial} - { m^*}} \right] 
\,\, \lambda_a i \gamma_5 \epsilon_p   =  0,
\end{equation}
\begin{equation}
\label{gap3}
\frac{\partial {\cal S}_{eff}}{\partial \phi_a}  \Big|_{\phi_a= <\phi_a>}  = 0 \, \to \,  
<\phi_a> = -  \left[ \mbox{Tr'}  \frac{i}{ i \slashed{\partial} - { m^*}} \right] \,\, \alpha
\left( T_a \frac{\varphi \cdot \lambda }{ \gamma \epsilon_s} +  \frac{3}{2 \alpha} \epsilon_3 \lambda_a \right) ,
\end{equation}
\begin{equation}
\label{gap5}
\frac{\partial {\cal S}_{eff}}{\partial \omega_a}  \Big|_{\omega_a= <\omega_a>} = 0  
\, \to \,   
<\omega_a> =  - \left[ \mbox{Tr'}  \frac{i}{ i \slashed{\partial} - { m^*}} \right] \,\,  \delta
 \left(   T_a  \frac{\lambda \cdot \varphi}{ \epsilon_s}  \right),
\end{equation}
where  the trace  Tr' is to be taken in all discrete indices (except flavor) and 
spatial integration.
In these equations  the effective mass matrix was defined as:
\begin{eqnarray}  \label{meff}
m^* =
m_0 - 
\frac{\epsilon_s  <\varphi_a > \lambda_a}{\beta} 
- \frac{3 \epsilon_3  <\phi_a>  \lambda_a}{2  \alpha} 
.
\end{eqnarray}
where the fields in brackets are the vacuum expectation values  and we have used  the following relations 
$$
\frac{\partial \epsilon_s}{\partial \phi_a} = \frac{ \beta T_a}{ \gamma   \epsilon_s} 
$$
and 
$$
\frac{\partial \epsilon_s}{\partial \omega_a} = \frac{ \beta T_a }{ \delta \epsilon_s}
$$
We next define the quantity $J$, common to all gap equations:
\begin{equation}
J = - \mbox{Tr'}  \frac{i}{ i \slashed{\partial} - { m^*}}
\label{defi}
\end{equation}
Since the above expression is  the trace of the Fourier transform of the quark propagator, we conclude that the effective mass
$m^*$ matrix must be diagonal in flavor space.  Therefore only $\phi_a$ with $a=3,8$ will contribute. 
The relevant gap  equations  can then be written as:
\begin{eqnarray} 
\label{gap-chiral-su3-2}
<\varphi_a> &=&   J  \epsilon_s \;  \lambda_a
\\
\phi &=&  \alpha  J \left(\frac{\varphi \cdot \lambda}{ 2 \alpha \epsilon_s  } 
+\frac{\epsilon_3 T \cdot \lambda}{\gamma T^2} 
\right) ,
\\
\omega &=&  \delta  J   \left( \frac{\varphi \cdot \lambda}{ \delta \epsilon_s } \right) .
\end{eqnarray}
where the conversions $<\phi_a> \rightarrow \phi$ and $<\omega_a> \rightarrow \omega$ were performed according to the 
prescription mentioned above and $\varphi \cdot \lambda = <\varphi_a> \cdot \lambda_a$  .
Multiplying the first equation above by $\lambda_a$, solving it for $J$ and substituting it into the other two
we obtain:
\begin{equation} 
\label{relation1}
\phi =  \frac{3 \alpha}{32 \alpha \epsilon_s^2}  (\varphi \cdot \lambda)^2
+ \frac{3 \alpha\epsilon_3}{ 16 \alpha \epsilon_s}  \frac{(\varphi \cdot \lambda) (T \cdot \lambda)}{  T^2  }
\equiv < (\bpsi \lambda_e \psi)^2 > + \epsilon_3  T_e \cdot < (\bpsi \lambda_e \psi) >  ,
\end{equation}
\begin{equation}
\omega =  \frac{3}{16 \epsilon_s^2} (\varphi \cdot \lambda)^2  
\label{relation4}
\end{equation}
where we have used the SU(3) relation: $ \lambda_a \lambda_a = \frac{16}{3}$.
And finally, using (\ref{omegapuro}), (\ref{phipuro}) and $ <\varphi_a> = \epsilon_s <\bpsi \lambda_a \psi>$   on  the above equations we find:   
\begin{equation} 
\label{relation2} 
< (\bpsi \lambda_e \psi)^2 >   =  \frac{3}{32}  (<\bpsi \lambda_a \psi>   \lambda_a)^2   
+ \epsilon_3  \left(  \frac{ 3 (T \cdot \lambda)}{ 16  T^2  } \lambda_a -  T_a  \right)  <\bpsi \lambda_a \psi> , 
\end{equation}
\begin{equation} 
< (\bpsi \lambda_d \psi )^2 >  +   < (\bpsi \lambda_d i \gamma_5 \psi)^2 >   =  
\frac{3}{16 } (<\bpsi \lambda_a \psi>   \cdot \lambda_a)^2
\label{relation3}   
\end{equation} 
As it can be seen, in (\ref{relation2}) factorization is violated by a constant factor (in front of the first term of the r.h.s)
and, more seriously, by the presence of the term linear in $<\bpsi \lambda_a \psi>$, which in turn comes from any non-vanishing value of 
$\epsilon_3$. From Eqs. (\ref{A-2}, \ref{A-3}, \ref{B-1}, \ref{B-2}, \ref{C-1}, \ref{D-1}, \ref{D-2}, \ref{E-2},  \ref{E-1}) 
we can see that  $\epsilon_3$ is determined mostly by $\kappa$, the coupling constant of the 't Hooft interaction.
Expression (\ref{relation3}) shows that a particular combination  of the two fourth order condensates does exhibits approximate factorization.
Moreover this expression suggests the existence of a fourth order chiral radius analogous 
to the usual one (i.e. $\bpsi \lambda_a \psi + \bpsi \lambda_a \gamma_5 \psi $).

It might seem that there is a large arbitrariness in the choice of  the shifts made above. 
However each one of them appear due to a precise reason, which is directly related to 
each of the terms of the original interactions.
There must be a shift on the composite $(\bpsi \lambda_d \psi )^2 +
(\bpsi \lambda_d i \gamma_5 \psi)^2$ because of the  $g_1$ interaction.
This was implemented in the shift of $\omega_a$ which 
couples to those fourth order terms.
On the other hand, if this shift includes any bilinear $\bpsi \Gamma \psi$
it  generates a sixth order term that does not exist in the original Lagrangian.
Therefore this bilinear cannot couple directly to the field $\omega_a$. 
This is the reason for the factorization of the condensate associated to $\omega$.
On the other hand the coupling of a bilinear is important for the cancellation of the 't Hooft term, 
and hence it must couple to the other field $\phi_a$.
This is the role of the term proportional to $\epsilon_3$,
which turns out to be  proportional to the parameter 
$\epsilon_5$, and both of them  proportional  to the 't Hooft term coupling $\kappa$ to ensure 
the corresponding cancellations.

\section{Discussion}

At this point it is interesting to remember some previous results  concerning the renormalization and renormalization scale 
dependence of the condensates discussed above \cite{gomez,narison-2}. 
In more phenomenological approaches, such as QCDSR, the condensates are numbers assumed to be finite and extracted from the analysis 
of data on hadron spectroscopy. They are not calculated and hence nothing can be  said about  possible divergences and their 
renormalization. In the NJL approach described above, the $<\bar{\psi} \psi>$ condensate can be calculated with the gap equation 
(\ref{gapsig}).  This calculation involves a UV divergent integral in momentum space, which represents a loop. Usually, at this point 
one introduces a cut-off which not only regularizes the integral but also sets a maximum momentum scale ($\Lambda$). The physics of the 
condensate is then "soft physics", involving only the low momentum quarks of the Dirac sea ($p < \Lambda$). In principle, 
although the models ($U(1)$ and $SU(3)$) considered here are non renormalizable,  we could try to write renormalization group equations 
to study the dependence of the results on the introduced scale $\Lambda$. However, since the number of parameters is very small and since 
we restrict ourselves to the description of low energy phenomena, it is enough to fit some data 
and fix the value of $\Lambda$. In first principle calculations of  $<\bar{\psi} \psi>$, such as lattice QCD, one has to compute 
the same loop taking into  account all the interactions described by the theory. As in the NJL case, one has to exclude the high 
momentum modes, i.e. the "perturbative part" of the condensate. This introduces a separation scale, which brings up the problems 
discussed above.

To summarize, in this work the auxiliary variable method was considered to investigate
the factorization of the four quark condensates $ <(\bar{\psi} \lambda_a  \psi)^2>$
and  $ <(\bar{\psi} \lambda_a    \gamma_5 \psi)^2>$
 into the quark-antiquark condensate
from a dynamical point of view with the SU(3) NJL effective model of QCD with 
higher order interactions.
For that an extension of the auxiliary variable method was applied to account for the several types of couplings of 
 multiquark  effective   interaction terms up to 
the eighth order.
Starting with  the one flavor U(1) invariant model it was possible to 
obtain   factorization. The peculiarity of this model is that it has no sixth order interaction.
For the case of the full SU(3) model, the method was applied with  
new auxiliary variables with their corresponding shifts.
 These shifts  are not very intuitive but they are 
needed to render possible all the cancellations of the  sixth and eighth order terms. 
Different shifts might be associated to different combinations of the two
four-quark condensates, such as those for the fields $\phi, \omega$, however 
the final result cannot present factorization for none of them.
In particular, the shift in the variable $\xi_a$, in expression (\ref{shift-2nd}), is needed to cancel the 't Hooft term and part of 
the eighth order interaction. Due to this shift, new bilinear terms appear in (\ref{relation2}). 
Therefore, in the case of the $ <(\bar{\psi} \lambda_e  \psi)^2>$ condensate,  the deviation from the factorization hypothesis 
can be traced back to the SU(3)  sixth order 't Hooft term. 
The inclusion of  a generic sixth order term  in the U(1) model, 
however,  would not change the results, i.e.
factorization would still appears in such flavorless model.
On the other hand, the ``chiral combination''  
$< (\bpsi \lambda_d \psi )^2 >  +   < (\bpsi \lambda_d i \gamma_5 \psi)^2 >$ can still be factorized, up to a constant factor.

\section*{Acknowledgements}

This work has been partially supported by FAPESP and CNPq, Brazil. We thank  J.R. Pel\'aez and J. Ruiz de Elvira for instructive discussions. 
F.L.B. is grateful for the hospitality extended to him at the IFUSP.

%% References with BibTeX database:

%\bibliographystyle{elsarticle-num}
%
%\bibliography{<your-bib-database>}

%{\bf 
%
%
%Maybe needless to do so:
%
%maybe one can find the gap equation , the complete one in adjoint rep,
%by defining:
%\begin{eqnarray}
%\frac{1}{i \slashed{\partial} - m - M_b \lambda_b} = A + B_b \lambda_b
%\\
%\mbox{such that one can find A and $B_b$ by means of}
%\\
% (A+ B_b \lambda_b )^{ij} ( i \slashed{\partial} - m - M_c \lambda_c )^{jk} = \delta_{ik}
%\end{eqnarray}
%With this representation 
%it would be easily to rewrite the gap equations with due care of the 
%$\lambda$ matrices structures..
%
%
%}

\end{document}